# DisSim-FinBERT[1]: Text Simplification for Core Message Extraction in Complex Financial Texts

Wonseong Kim[1] **, Christina Niklaus[2], Choong Lyol Lee[3], Siegfried Handschuh[2]

[1] Advancing Systems Analysis Program, International Institute for Applied Systems Analysis (IIASA), Laxenburg, Austria
[2] ICS, University of St. Gallen (HSG), St. Gallen, Switzerland
[3] Department of Economics and Statistics, Korea University, Sejong, Republic of Korea

** Corresponding author: E-mail: kimwonseong@iiasa.ac.at

**Abstract**

This study proposes DisSim-FinBERT, a novel framework that integrates Discourse Simplification (DisSim) with Aspect-Based Sentiment Analysis (ABSA) to enhance sentiment prediction in complex financial texts. By simplifying intricate documents such as Federal Open Market Committee (FOMC) minutes, DisSim improves the precision of aspect identification, resulting in sentiment predictions that align more closely with economic events. The model preserves the original informational content and captures the inherent volatility of financial language, offering a more nuanced and accurate interpretation of long-form financial communications. This approach provides a practical tool for policymakers and analysts aiming to extract actionable insights from central bank narratives and other detailed economic documents.

---

[1] The implementation of DisSim-FinBERT-FOMC is publicly available at https://huggingface.co/Wonseong/dissim-finbert-fomc.

# 1 Introduction

Effective communication by central banks has become a strategic part of monetary policy. This is due to its power to shape market expectations, a growing public demand for clear information about bank actions, and its role in improving policy effectiveness (Casiraghi & Perez, 2022). Furthermore, open and clear communication supports the bank's accountability and independence. IMF (Casiraghi & Perez, 2022) published 'Monetary and Capital Markets Technical Assistance Handbook for Central Bank Communication' and indicates five effective communication guidelines: Transparency, Reach, Regularity, Accessibility, and Consistency. In line with this trend, the Federal Reserve Board (FRB) aligns 'Press Release' as textual data on website[2].

The Federal Open Market Committee (FOMC) is crucial in guiding US monetary policy and, indirectly, the world economy. As the main decision-making entity of the Federal Reserve System, the FOMC's communication approaches significantly influence financial markets and economic stability. The FOMC holds official meetings eight times per year. Approximately 20 days after each meeting, materials known as the 'Minutes of the Federal Open Market Committee' become accessible. The Minutes include the current financial situation, future economic outlook, and committee policy actions. An in-depth study of these Minutes can provide efficient access to expert group analysis and discussions.

However, a cautious approach is necessary when analyzing the Minutes using language models. The Minutes are not just a few key sentences like news reports; they consist of numerous lengthy paragraphs. Deciphering their message can be timeconsuming, even for experts, given the complexity of financial texts. Moreover, technical and dense language can be a hurdle for individuals without an extensive background in economics or finance, making it difficult for the general public to understand the content. In the context of the Minutes of FOMC, the presence of multiple aspects within a single sentence can pose challenges on large language models, such as BERT and GPT4. For instance, a sentence may discuss both inflation and unemployment, making it difficult for the model to isolate and understand each aspect individually (Kim, 2023). Mitigating this limitation may involve re-training the model on a corpus that includes more complex sentence structures and a broader range of aspects. However, such an approach would require significant computational resources and domain expertise.

[2] Board of Governors of the Federal Reserve System (https://www.federalreserve.gov/)

Furthermore, the performance limitations of FinBERT could be attributed to the training data used. The model's performance may be constrained if the training data does not cover all aspects of the financial sector. This paper proposes a solution to handle complex financial text using discourse simplification (Kim & Lee, 2024) created by a coauthor. We aim to enhance ABSA's performance in complex financial text and Fed communication by breaking down the complex high-level text into shorter canonical structures.

## 2 Related Work

Recent studies have proposed innovative approaches to improve sentiment analysis and information retrieval in domain-specific contexts. Jin (Jin et al., 2023) introduced the Fintech Key-Phrase dataset, a large-scale, human-annotated resource tailored for expression-level information retrieval in the Chinese financial high-tech sector. By combining domain expertise with data augmentation techniques such as ChatGPT, this dataset significantly improves the accuracy of key-phrase extraction, highlighting the importance of specialized datasets for financial NLP tasks. Complementing this, WordTransABSA (Jin et al., 2024) proposed a novel method for enhancing aspect-based sentiment analysis (ABSA) by leveraging the pre-training paradigm of masked language modeling. Instead of discarding the decoder, their *Word Transferred LM* strategy uses affective token prediction to better utilize the full capacity of pre-trained language models. These works collectively underscore the growing relevance of domain-specific modeling and pre-training-informed architectures, aligning with our motivation to improve sentiment understanding in complex financial texts through simplified and structured preprocessing.

Lesmy et al. (Lesmy et al., 2024) show that increasing intricacies in financial disclosures hinder readability, affecting stakeholders' ability to interpret key information. This aligns with the focus of this paper, which examines how the clarity of financial texts, as reflected in sentiment analysis, influences economic decision-making. Levy et al. (Levy et al., 2022) and Blum and Raviv (Blum & Raviv, 2023) emphasize the role of crises and regulations in shaping financial communication. Levy et al. observed delayed recognition of the 2008 crisis by economists, while Blum and Raviv analyzed shifts in banks' risk disclosures post-Basel III. These works underscore the significance of transparent financial communication, a theme that complements this study's exploration of sentiment-driven market expectations.

The current state of research involves the use of metadata or word-level analysis for data generation in economic analysis on the behavior of market agents (Hansen et al., 2018; Kim, 2023; Rosa, 2013; Stekler & Symington, 2016; Tadle, 2022). While this intuitive approach may offer some valuable insights, it may not fully capture the precise intentions of central bankers. Consequently, there is a growing interest in exploring the potential of neural network language models, which can enable more advanced and nuanced text analysis in this domain. Among numerous machine learning models used in Natural Language Processing (NLP), the Bidirectional Encoder Representations from Transformers (BERT) (Devlin et al., 2019) model has gained prominence for its ability to understand sentence context bidirectionally, surpassing previous models using Recursive Neural Network (RNN) models. RNNs are traditionally trained to predict either the succeeding word in a sentence (utilizing the context on the right side) or the preceding word, but not both simultaneously.

However, BERT's broad focus can limit its performance in specialized domains, where central bankers use complex sentences in communications. To address this challenge, FinBERT (Araci, 2019) has been introduced, which is a domain-specific version of BERT pre-trained on a large-scale financial text corpus. FinBERT has demonstrated significant improvements in accuracy for sentiment analysis and other NLP tasks within the financial domain (Kim & Lee, 2024). In a recent study, Gössi et al.(2023) conducted a study in 2023 where they fine-tuned the FinBERT model using a dataset derived from the Federal Open Market Committee (FOMC) minutes. The dataset was manually labeled with sentiment indicators. This approach involved training on complex sentences and resulted in improved performance compared to the original Fin-BERT model, which was manually labeled with sentiment indicators. This approach, which involved training on complex sentences, demonstrated an enhanced performance compared to the original FinBERT model (Kim et al., 2024).

Aspect-based sentiment analysis (ABSA) is an extension of traditional sentiment analysis that determines sentiment polarity and identifies specific aspects or entities associated with the sentiment. ABSA aims to discern sentiment polarity towards specific aspects within a given context, providing more precise insights from textual data. In recent years, ABSA has gained interest in analyzing the Federal Open Market Committee (FOMC) minutes (Wang, 2021), which contain crucial insights about the U.S. economic outlook and monetary policy decisions. Given the significant impact of the central bank's communication on financial markets, ABSA can offer a nuanced understanding of the sentiment conveyed in the FOMC minutes, benefiting financial institutions, policymakers, and investors.

On the other hand, applying ABSA to FOMC minutes is challenging due to the complexity of the text and its specific context (Kim et al., 2023). The communication from central banks is characteristically neutral, maintaining a careful balance to avoid undue influence on the financial market. This impartiality is crucial as every phrase, sentence, or statement could have profound implications for the market dynamics. Furthermore, the communication often comprises complex and nuanced statements that concurrently address multiple aspects. In Wang's (Wang, 2021) aspect-based sentiment analysis (ABSA), they used cosine similarity to identify aspects in the text. This method involved calculating tokenized sentences based on pre-identified aspects such as inflation, employment, and investment. However, the similarity values obtained were often very close to each other, which could lead to uncertainties in aspect selection. To address this complexity, our approach suggests using grammatical simplification through discourse simplification.

## 3 Dataset

The dataset comprises 32,034 sentences of FOMC communications ranging from January 3, 2006, to February 22, 2023. These communications have been extracted from the minutes of the FOMC meetings. Out of the total 32 thousand sentences, 1,030 have been manually labeled based on their sentiment regarding economic growth, employment, and inflation(Kim et al., 2023). The dataset has been labeled by three researchers[3] in the lab. It was decided that a minimum of two agreements was needed for each aspect. During the labeling process, one of the researchers was unaware of the other researchers' labeling results. If all three researchers labeled different aspects, each positive, neutral, and negative, then it was decided that 'No Majority Found,' and the data was not included in the dataset.

# 4 Solution: Extract Core Message in Communication

DisSim, short for "Discourse Simplification," refers to a technique used to simplify complex text by breaking it down into shorter sentences with a canonical structure (Niklaus et al., 2019). The primary objective of discourse simplification is to enhance understanding, especially for texts that feature complex language structures, technical terminology, or domain-specific comprehension. Through discourse simplification, the information within the text is decomposed into lower-level sentences, allowing for easier comprehension and interpretation.

## 4.1 Approach

DisSim is a discourse-aware sentence-splitting approach for English that creates a semantic hierarchy of simplified sentences. It takes a sentence as input and performs a recursive transformation process that is based upon a small set of 35 hand-crafted grammar rules. These patterns were heuristically determined in a comprehensive linguistic analysis and encoded syntactic and lexical features that can be derived from a sentence's parse tree. Each rule specifies (1) how to split up and rephrase the input into structurally simplified sentences and (2) how to set up a semantic hierarchy between them. They are recursively applied to a given source sentence in a top-down fashion. When no more rule matches, the algorithm stops and returns the generated discourse tree[4].

---

[3] Three postgraduate researchers labeled the dataset using a majority voting system. In the first round, two researchers (label-1 and label-2) each labeled 1,500 sentences, providing two initial labels for each sentence. In the second round, another researcher (label-3) labeled sentences inconsistently classified in the first round, contributing to a third label. Sentences not classified by at least two of the three labelers were excluded, yielding a final set of 1,030 labeled sentences after removing 321 duplicates. All labelers were postgraduate students with expertise in business, data science, and finance, with the third labeler having a BA and MSc in financial economics.

[4] The source code of framework is available under https://github.com/Lambda-3/DiscourseSimplification

### *Split into Minimal Propositions*

In the first step, source sentences that present a complex linguistic form are turned into canonical structures by decomposing clausal and phrasal components. For this purpose, the transformation rules encode both the splitting points and the rephrasing procedure for reconstructing grammatically sound sentences.

### *Establish a Semantic Hierarchy*

To create a semantic hierarchy among the split sentences, two sub-tasks are performed:

(i) **Classification of Constituency Types:** Initially, a contextual hierarchy is established by linking the split sentences with information about their hierarchical level, similar to the concept of nuclearity in Rhetorical Structure Theory (Mann & Thompson, 1988). The core sentences (nuclei) contain the main information of the input, while the accompanying contextual sentences (satellites) provide additional details. To differentiate between these two types of constituents, a straightforward syntax-based approach is employed in the transformation patterns. Subordinate clauses/phrases are classified as context sentences, while superordinate and coordinate clauses/phrases are labeled as core.

(ii) **Identification of Rhetorical Relations:** The objective is to restore the semantic relationship between the disembedded components. To achieve this, the rhetorical relations existing between the simplified sentences are identified and classified. This process utilizes both syntactic features derived from the parse tree structure of the input and lexical features in the form of cue phrases. Inspired by the previous work (Taboada & Das, 2013), a predefined list of rhetorical cue words is employed to infer the type of rhetorical relation.

## 4.2 Advantages

DisSim presents two primary advantages when applied to FOMC minutes, particularly for extracting accurate sentiment predictions from intricate financial text.

#### *Finding a key statement*

FinBERT's classification performance can degrade due to cosine similarity across various aspects, often missing key sentences. To compensate, sentiment analysis is sometimes performed by considering nearby sentences as key statements. As illustrated in Figure 1, DisSim decomposes sentences into several levels, where level 0 designates the key statement in the text. This methodology can alleviate the ambiguity often encountered in text classification and enhance the accuracy of sentiment prediction.

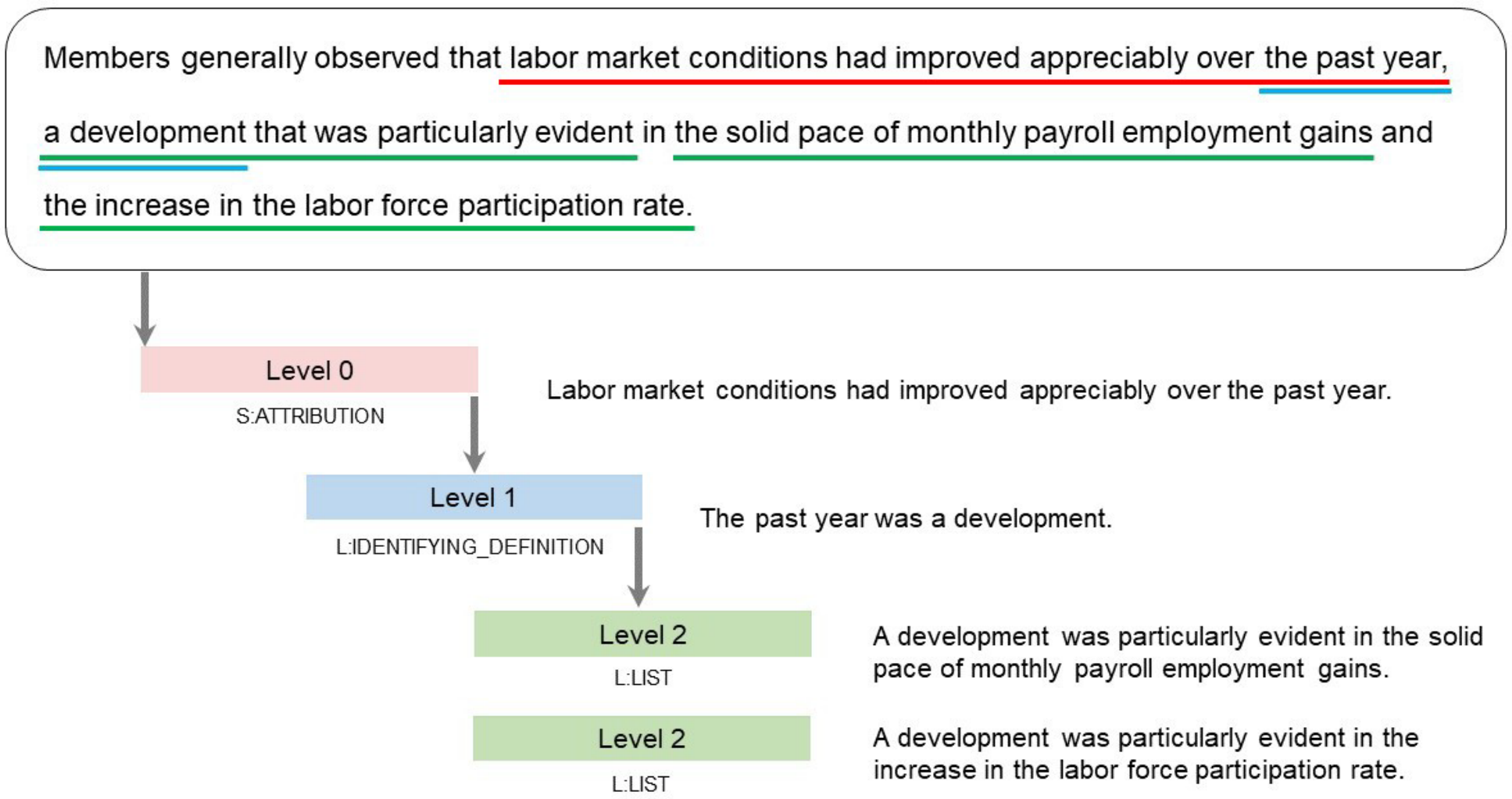

**Fig. 1 Hierarchical Discourse Simplification (DisSim) process.** The diagram illustrates how sentences are decomposed into multiple levels of significance, with Level 0 representing the key statement (labor market conditions improving), Level 1 containing identifying definitions, and Level 2 listing supporting evidence (monthly payroll gains and labor force participation rate). This decomposition helps address cosine similarity issues in FinBERT classification by clearly identifying the primary sentiment-bearing statements.

### *Hierarchical Segmentation*

In the Figure 1, each level are allocated in a hierarchical structure or rhetorical relation. Level 1 as a subordinate clause of Level 0 explains 'a development has linked in the past year'. Level 2s are listed as 'a development' which leads two parallel phrases. In the example sentence, subordinate clauses take a role as supporter of key statement (Level 0), while some are contrasted with them and change the sentiment in opposite direction. Be Careful on applying DisSim on complex sentences. In this paper, not considering the contrasted part, only Level 0 sentences are implicated to enhance ABSA by FinBERT.

As depicted in Figure 1, each level in DisSim is arranged in a hierarchical structure linked via rhetorical relations. Level 1, serving as a subordinate clause of Level 0, describes that a development has been linked in the past year. Level 2 elements list details about 'a development', leading to two parallel phrases. In the given example sentence, subordinate clauses often support the key statement (Level 0), but some may contrast and change the sentiment in the opposite direction. It's essential to be mindful of this complexity when applying DisSim to complex sentences. For the purpose of this paper, we focus only on 'Level 0' sentences, as the core messages of FOMC minutes, to test the enhancement of Aspect-Based Sentiment Analysis (ABSA) by FinBERT (Gössi et al., 2023).

# 5 New Model: DisSim-FinBERT

We propose incorporating a more detailed preparatory stage for Aspect-Based Sentiment Analysis (ABSA). In Wang's model (Wang, 2021), which serves as the baseline, FinBERT was used to classify FOMC minutes based on cosine similarity. However, the model struggled to accurately identify multiple aspects, resulting in the misclassification and incorrect allocation of aspects in the prediction outcomes.

Figure 2 illustrates the workflow of our new model, which incorporates the DisSim method before aspect selection using cosine similarity. The process begins by inputting complex sentences into the DisSim filter, which extracts 'Level 0' key statements. Selecting key sentences might lead to a loss of information from the complete sentence. Therefore, Discourse simplification is suggested only during the aspect selection stage. For the classification of economic sentiment in FOMC minutes, we utilize the full sentence without simplification to ensure comprehensive analysis.

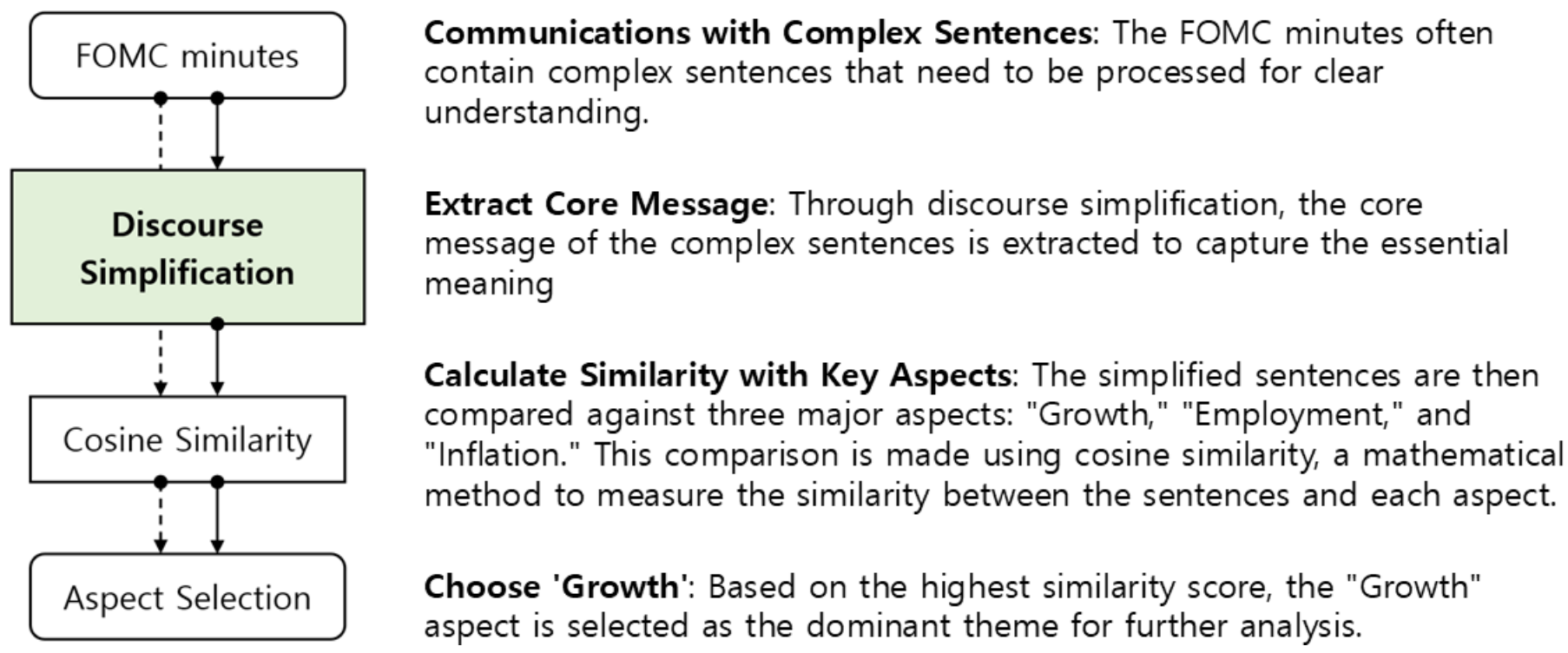

**Fig. 2 Process of Aspect-Based Sentiment Analysis.** The flowchart illustrates the pipeline for processing FOMC minutes, beginning with discourse simplification to extract core messages from complex sentences, followed by cosine similarity calculations to measure relevance to key economic aspects ('Growth,' 'Employment,' and 'Inflation'), and concluding with aspect selection based on the highest similarity score. The diagram demonstrates how this methodology enhances the baseline FinBERT model's performance by addressing the challenges of complex financial text.

## 5.1 Aspect Selection Performance

'Level 0' statements are then evaluated using cosine similarity to determine their corresponding aspects. We categorize aspects into 'growth', 'employment', and 'inflation'. Table 1 showcases the shift in the number of sentences according to these aspects. This model, therefore, offers a systematic approach for dissecting complex sentences and aligning them with their appropriate aspects, resulting in more accurate sentiment predictions. To evaluate the effectiveness of the new model, we utilize the labeled dataset from the FOMC (Kim et al., 2023). This dataset encompasses 1,030 labeled sentences, which are categorized under three distinct aspects: growth, employment, and inflation.

### *Aspect Selection Test*

Initially, the cosine similarity for each of the 1,030 sentences is calculated, and the aspect with the highest value is selected as the main aspect. The results of this process are presented in Table 1.

The FinBERT model predominantly categorizes the main aspect as 'growth', as indicated by 981 out of the 1,030 sentences. In contrast, the DisSim-FinBERT model diversifies the aspect allocation, not solely focusing on ’growth’.

**Table 1 Results of Aspect Selection**

| Aspect | FinBERT | DisSim-FinBERT |
|---|---|---|
| growth | 981 | 842 |
| employment | 16 | 28 |
| inflation | 33 | 160 |
| total | 1,030 | 1,030 |

*Note: This table compares the aspect distribution between the baseline FinBERT model and the enhanced DisSim-FinBERT approach across 1,030 analyzed sentences. The FinBERT model shows a strong bias toward the growth aspect (981 sentences, 95.2%), whereas DisSim-FinBERT demonstrates a more balanced distribution with notable increases in inflation-related classifications (from 33 to 160 sentences) and employment-related classifications (from 16 to 28 sentences).

For instance, consider the original sentence: ’Inflation pressures in foreign economies generally remained subdued, even though higher oil prices put some upward pressure on headline inflation.’ Here, the FinBERT model assigns it to the ’growth’ aspect. However, after the application of the DisSim method, the sentence transitions to the ’inflation’ aspect, represented by the statement: ’Inflation pressures in foreign economies generally remained subdued.’ This example demonstrates the ability of the DisSim-FinBERT model to more accurately classify complex sentences according to their primary aspect.

***Sentiment Prediction Test***

Clearly, different aspect selections reveal different sentiment predictions, especially in time series data. Figure 3 a comparative view of three bar charts, each quantifying aspect-based sentiment analysis results directed at ’Economic Growth’ and classified into negative, neutral, and positive sentiments. Each chart is labeled as follows: Chart A, human label, termed ’Sentiment Label,’ depicts negative sentiment in purple, neutral in green, and positive in blue. The green bar representing neutral sentiment stands out as the most prominent, reaching a value of 527. It is followed by positive sentiment at 262 and negative sentiment, the least, at 241. In contrast, Chart B, FinBERT with no sentence simplification, labeled 'FinBERT-FOMC,' demonstrates a leftward skew with the negative sentiment peaking at 415, overshadowing the neutral at 339 and positive at 276. Chart C, FinBERT-FOMC result with simplified sentence, mirrors a similar pattern to Chart A, with neutral sentiment once again leading at 435, followed by negative at 352 and positive at 243. This similarity indicates that DisSimFinBERT's simplified approach to aspect selection produces results comparable to those manually labeled by human analysts.

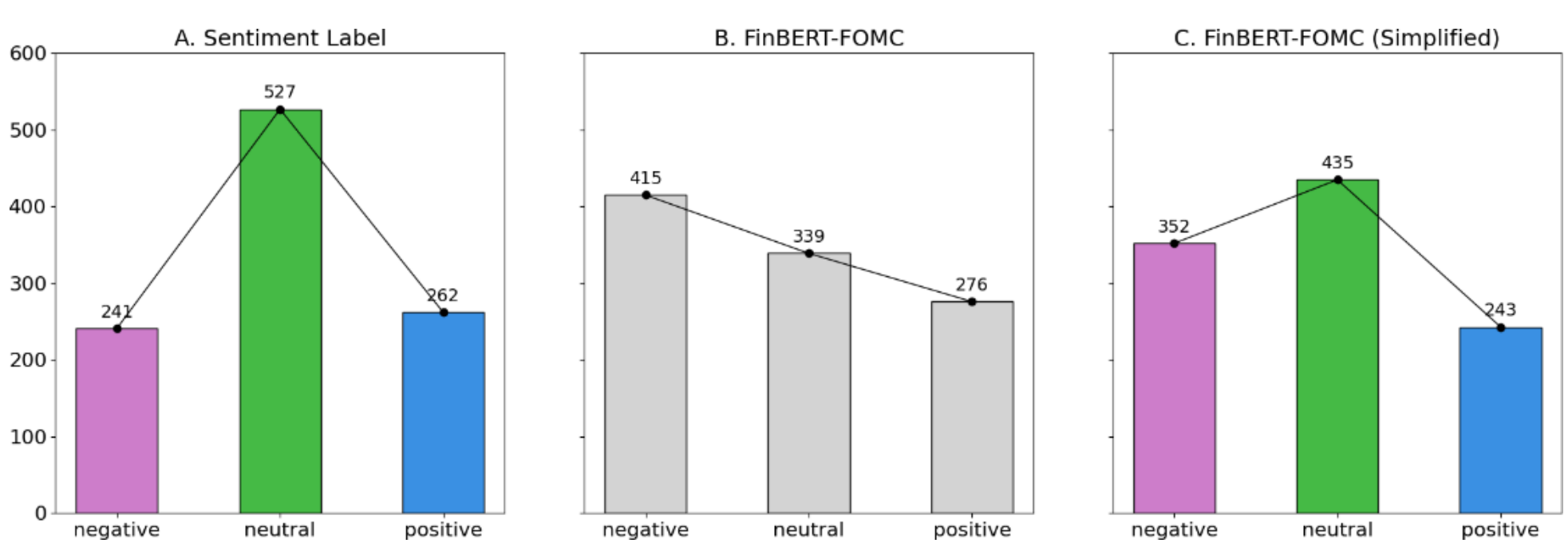

**Fig. 3 Comparative Aspect-Based Sentiment Analysis on Economic Growth.** The figure presents sentiment distributions across three methods: (A) Human-labeled sentiment (Sentiment Label), (B) FinBERT-FOMC without sentence simplification, and (C) DisSim-FinBERT with discourse simplification applied during aspect selection. Each bar chart shows the number of sentences classified as negative, neutral, or positive. Compared to FinBERT-FOMC, the DisSim-FinBERT model produces a sentiment distribution more aligned with human labels, notably increasing neutral classifications and reducing negative bias. This highlights the effectiveness of discourse simplification in improving sentiment accuracy in complex economic texts.

The analysis of Chart B indicates a potential bias in economic analysis as it shows negative sentiment predominantly at 415. This is in contrast to Chart A's manual labeling and Chart C's simplified model, which present neutral sentiment as the most prevalent. Such a difference in aspect selection could lead to an overly pessimistic interpretation of 'Economic Growth,' resulting in skewed analysis. This anomaly highlights the importance of carefully evaluating sentiment analysis models, especially when applied to complex economic texts, to ensure accurate and balanced economic insights.

## 5.2 Justifying Methods by Alignment with Real Economic Events

The Federal Open Market Committee (FOMC) releases minutes eight times a year, creating an irregular monthly pattern with four months lacking data annually. This inconsistency poses challenges when aligning FOMC data with other economic indicators, which are typically reported on a monthly, quarterly, or annual basis. Additionally, sentiment predictions derived from FOMC minutes tend to be more volatile compared to traditional economic indicators. To ensure meaningful integration of these data types for economic analysis, it is essential to smooth the FOMC sentiment data. This involves transforming the sentiment index into a more continuous format that mitigates extreme fluctuations in text valence, making it more compatible with macroeconomic analyses.

A few key steps must be followed to achieve a smoother integration of FOMC sentiment data with other economic indicators. First, the infrequent nature of FOMC communication must be addressed by applying a smoothing technique that interpolates or averages sentiment scores across the months where no data is released. This helps reduce the abrupt shifts caused by the irregular release schedule. Second, selecting an appropriate filter is crucial to maintaining the integrity of the sentiment data while minimizing noise. The filter should be carefully chosen to balance sensitivity to sentiment changes and the need for stability in the data. Lastly, it is important to align the smoothed FOMC sentiment data with major economic events, such as GDP releases, inflation reports, or employment data, to ensure that any sentiment shifts are meaningfully compared and contextualized within the broader macroeconomic landscape. This step enhances the relevance and interpretability of the analysis, allowing for a more cohesive understanding of the data.

### 5.2.1 Smoothing of High-Frequency Time-series Data

A moving average is a commonly used method for smoothing data, as described by the Federal Reserve Bank of Dallas[5]. This approach helps create a more continuous and stable time series by averaging data points over a specific period. However, despite its effectiveness in general smoothing, this method has notable limitations when applied to sentiment analysis. Specifically, in the context of a sentiment index, a moving average may oversimplify the data, smoothing out critical nuances and fluctuations essential for understanding underlying economic conditions or shifts in monetary policy. Therefore, applying a moving average to sentiment analysis requires careful consideration to avoid losing valuable information embedded in the sentiment data.

---

[5] Smoothing data with moving averages, https://www.dallasfed.org/research/basics/moving

Various techniques are available for smoothing high-frequency data, including the Moving Average, Hodrick-Prescott (HP) Filter, Wavelet Transformation, and the Savitzky-Golay (SG) Filter. Each method possesses unique characteristics and is suited for different types of data and analysis. The Hodrick-Prescott Filter, for instance, is frequently used in macroeconomic time series to separate cyclical fluctuations from long-term trends. Its effectiveness hinges on the smoothness parameter[6] (lambda, λ), which balances the trade-off between preserving cyclical information and smoothing the data (Hodrick & Prescott, 1997). However, it may introduce distortions, particularly at the sample's endpoints. The Wavelet Transformation (Daubechies, 1990), on the other hand, excels in handling non-stationary data by allowing multi-scale analysis, making it ideal for capturing both shortand long-term fluctuations. The Savitzky-Golay (SG) Filter (Hamming, 1998; Schafer, 2011), in contrast, maintains critical data features, such as local maxima, minima, and widths, by employing a polynomial smoothing technique that preserves significant aspects of the data.

When selecting a smoothing method for predicting economic fluctuations, it is essential to consider the specific characteristics of the data and the type of predictions required. Each method presents trade-offs between smoothness and fidelity to the original data, the importance of capturing short versus long-term trends, and the computational complexity involved. For instance, the best criterion for smoothing FOMC sentiment data is to capture the timing of economic upturns and downturns accurately. Well-filtered data can highlight periods of economic crisis or significant downturns, locally minimizing average sentiment when necessary.

Figure 4 presents four smoothed plots of FOMC sentiment data using different techniques. The dataset spans from 2006 to 2023, including key economic periods such as the Global Financial Crisis, the European Sovereign Debt Crisis, and the

[6] Typical values for lambda ($\lambda$) vary based on the data's frequency: annual=100, quarterly=1,600, monthly=14,400. However, there is no universally fixed $\lambda$ value.

COVID-19 pandemic. The visualizations demonstrate how each method reflects these significant events, underscoring the importance of selecting the most appropriate smoothing technique based on the nature of the data and the analysis objectives.

***Moving Average Smoothing***

The top plot shows the data smoothed using a moving average (window = 12 months), a technique that helps to identify trends by averaging data points over a specific time period. The sentiment fluctuates over time, with noticeable dips during the European Sovereign Debt Crisis and the Covid-19 Pandemic.

***HP Filter Smoothing***

The second plot from the top shows the data after applying the Hodrick-Prescott (HP) filter. This plot displays a clear cyclical pattern, with peaks and troughs corresponding to various economic events. Economic downturns occurred during the Global Financial Crisis, the European Sovereign Debt Crisis, and the Covid-19 Pandemic. Due to a large number of missing values in the original data, even if we use a lambda parameter of 10, the resulting plot will still be very smooth. However, it's possible that sentimental information may have been ignored.

***Wavelet Smoothing***

The third plot demonstrates the use of Wavelet[7] transformation for smoothing. The resulting plot appears to retain more of the high-frequency noise compared to the other methods. Moreover, it may not reveal economic downturns during a major crisis.

***SG Filter Smoothing***

The fourth plot in the figure shows the Savitzky-Golay7filter[8], which is an excellent tool for smoothing sentiment data. It can highlight crucial economic crises without losing significant features of sentiment trends, such as peaks and valleys that correspond to economic events. This filter helps reveal the nuances of the Global

[7] Hyper-parameters: type=haar, level=1, mode=per, threshold=0.9, threshold mode=soft

[8] Hyper-parameters: window length = 20, poly order = 2, deriv = 2, delta = 1, axis = 0, mode = wrap, cval = 0

Financial Crisis, the European Sovereign Debt Crisis, Stock Market Selloff, and the COVID-19 pandemic making sure that the significant shifts in sentiment associated with these events remain visible. The SG filter is the preferable choice for analyzing sentiment data, especially when distinguishing between major economic disruptions within a smoothed dataset, as it maintains the essential characteristics of the data.

### 5.2.2 Why the Savitzky-Golay (SG) Filter is Ideal for Sentiment Time Series Smoothing

The Savitzky-Golay (SG) filter is a powerful method for smoothing time-series data, particularly for sentiment analysis, due to its ability to preserve key data characteristics while minimizing noise. Below are several reasons why the SG filter is ideal for sentiment time-series data, accompanied by mathematical explanations.

***Preservation of Key Data Characteristics (Peaks, Troughs, Extremes)***

The SG filter works by fitting a polynomial to a local window of data points via leastsquares regression. Given a time-series data set y(t), the SG filter fits a polynomial P (t) of degree k to a set of 2m+1 points in the window around time t. Mathematically, this can be expressed as:

$$P(t) = a_0 + a_1 t + a_2 t^2 + \cdots + a_k t^k$$

The coefficients a0, a1, . . . , ak are chosen such that they minimize the squared error between the polynomial and the actual data points. This polynomial is then used to smooth the central point in the window, and the process is repeated for every point in the time series.

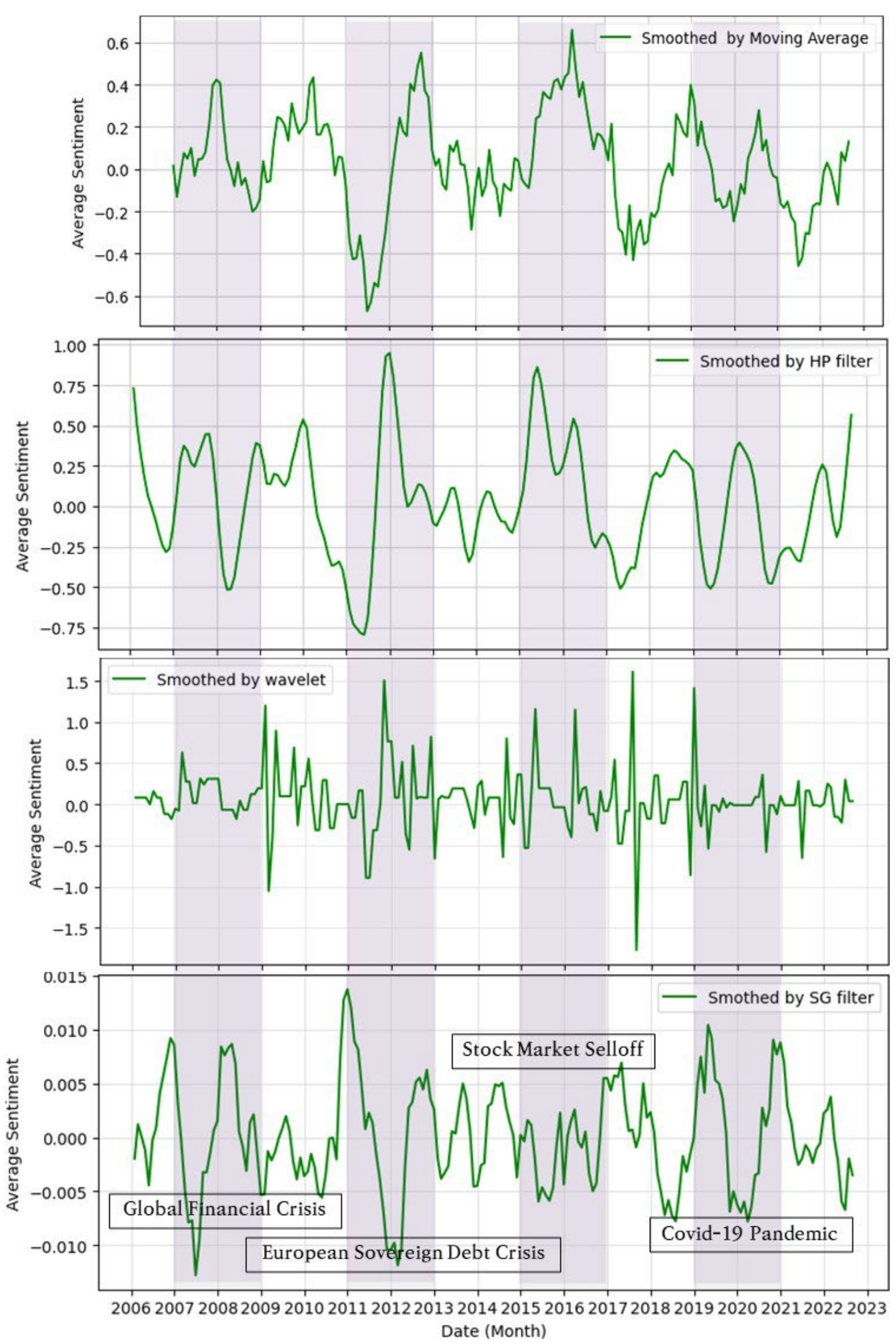

**Fig. 4  Smoothed Growth Sentiment Using Four Filtering Methods (2006–2023).** The figure compares Moving Average, HP Filter, Wavelet, and Savitzky-Golay (SG) smoothing techniques applied to FOMC-based growth sentiment. Shaded areas mark major economic crises. Among the methods, the SG filter best captures key downturns (e.g., 2008, 2012, 2015, 2020) while preserving sentiment structure and volatility.

Because the SG filter fits a polynomial, it preserves local features like peaks and troughs, which are often lost when using simpler smoothing techniques such as the moving average. The polynomial form allows the filter to follow the shape of the data more closely, capturing key shifts in sentiment.

***Reduction of Noise While Retaining Underlying Trends***

The SG filter effectively reduces high-frequency noise by smoothing over a local window of points. Since the filter applies least-squares polynomial fitting, it is able to remove random fluctuations that occur at higher frequencies, while still retaining the overall trend. The degree of the polynomial k and the window size 2m+1 control the trade-off between noise reduction and trend retention.

By setting an appropriate degree of the polynomial, the SG filter avoids oversmoothing, which is common with simpler methods like the moving average. Oversmoothing can cause significant deviations from the true underlying trend, but the polynomial form of the SG filter retains the shape of the data, ensuring that important trend changes are not lost.

***Flexible Smoothing Options***

The SG filter allows flexibility through the choice of polynomial degree k and window size 2m + 1. For example, if you use a larger window or a higher-order polynomial, the filter will smooth more aggressively, reducing noise but possibly sacrificing some fine detail. On the other hand, a smaller window or lower-order polynomial will better retain local variations in sentiment but may leave more noise in the data.

The choice of k and 2m + 1 is guided by the specific nature of the time-series data. For sentiment analysis, where both long-term trends and short-term fluctuations matter, this flexibility ensures that the SG filter can be fine-tuned to optimally smooth the data without losing critical sentiment information.

***Suitability for Nonlinear Sentiment Data***

Sentiment data is often nonlinear, and the SG filter is well-suited to handle such nonlinearity because it uses polynomial fitting, which can approximate both linear and nonlinear trends locally. The local polynomial approximation handles small-scale nonlinearity better than techniques like moving averages, which assume linearity over a window of points.

This local polynomial fitting is particularly important for sentiment analysis, where sudden changes in sentiment might not follow a linear trend. The SG filter's ability to capture these shifts ensures that nonlinear fluctuations in sentiment data are not lost, providing a more accurate representation of how public sentiment responds to economic or political events.

***Effectiveness in Capturing Economic Events***

In the context of analyzing FOMC sentiment or other economic time series, the SG filter is particularly effective at highlighting periods of economic crisis. For example, during the Global Financial Crisis or the Covid-19 pandemic, sentiment shifts rapidly, leading to sharp peaks or troughs in the data. The SG filter's polynomial fitting ensures that these critical points are preserved, as opposed to being smoothed over by more basic filters.

The mathematical robustness of the SG filter allows it to retain these sharp transitions, which are often key to understanding public sentiment during turbulent economic times. While other filters, like the moving average, may smooth out these important shifts, the SG filter preserves them by fitting the polynomial more closely to the original data.

### 5.2.3 Graphical Representation by SG filter

We sought to identify empirical evidence using ABSA results, with DisSim-FinBERT employed for aspect selection. Figure 5 presents three time-series graphs titled "Growth Sentiment by Label," "Growth Sentiment by FinBERT-FOMC," and "Growth Sentiment by FinBERT-FOMC (Simplified)." Each graph displays sentiment scores ranging from -1 to 1, where -1 represents strong negative sentiment, 0 indicates neutrality, and 1 denotes strong positive sentiment. The dataset follows a daily frequency based on the dates published on the Federal Reserve's website. As a result, time-series data was subject to volatility, and the typically available macroeconomic indicators (monthly, quarterly, or annually) were omitted. The daily sentiment data required pre-processing to align the time steps between the two datasets for economic analysis. This transformation ensured that the data was continuous and less reactive to extreme valence fluctuations in the text. The SG filter was applied to smooth the data by fitting a polynomial to the points, effectively reducing noise while preserving the shape and amplitude of the signal peaks. This approach is beneficial for analyzing data where peak shapes carry critical information.

***Growth Sentiment by Label***

The graph depicts sentiment over time, represented by a black line, with data smoothed using a Savitzky-Golay filter. A red dashed line illustrates the moving average for comparison. The sentiment exhibits notable fluctuations, frequently crossing the zero line. Our research hypothesis suggests that, when FinBERT-FOMC is applied with appropriate simplifications, it will align closely with the labeled results and demonstrate an improved ability to capture economic fluctuations, particularly during periods of turbulence.

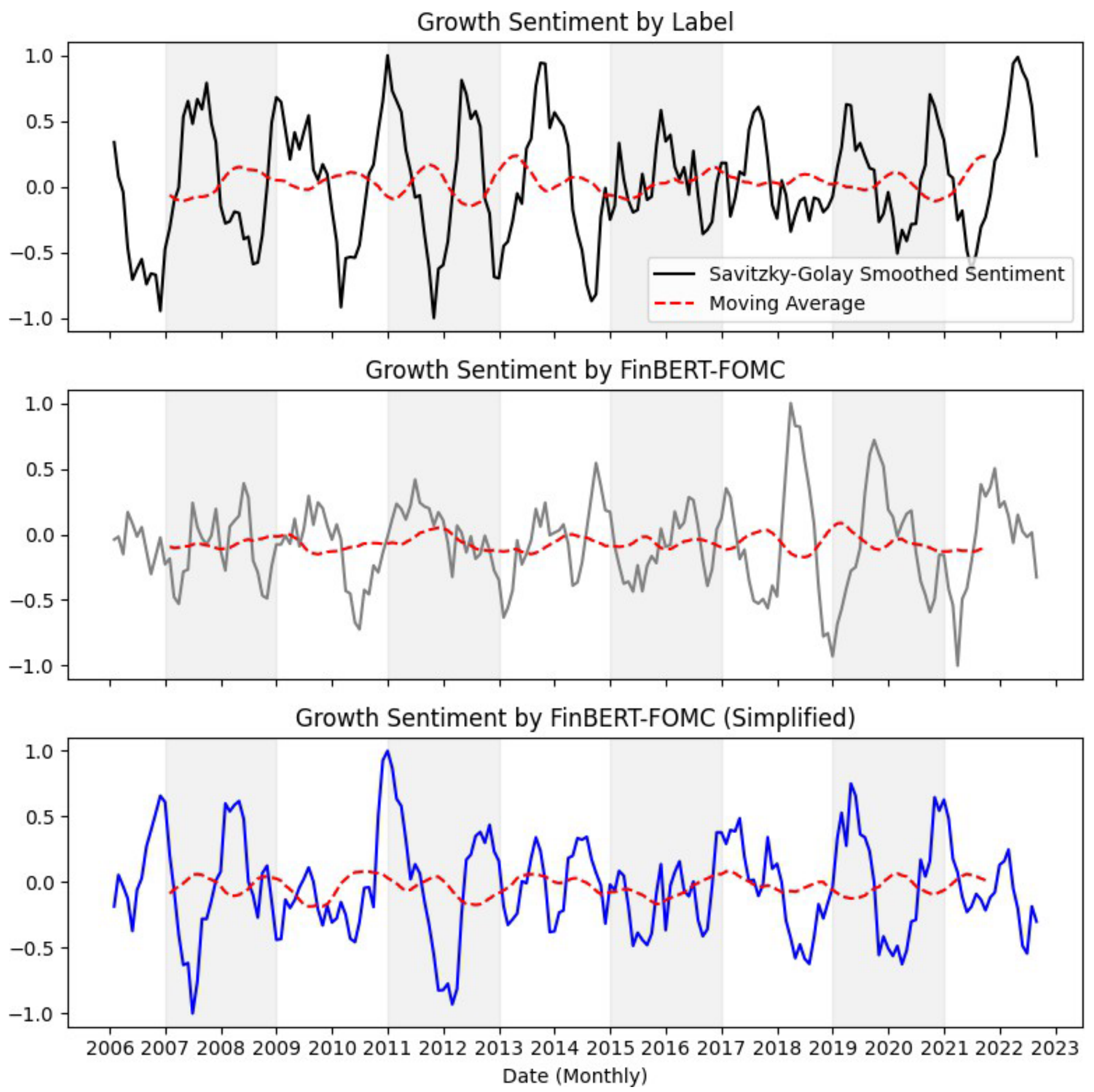

**Fig. 5 Smoothed Time-Series of Growth Sentiment from Three Sources.** The panels compare growth sentiment over time using (top) human-labeled data, (middle) FinBERT-FOMC, and (bottom) DisSim-FinBERT (Simplified). Each includes Savitzky-Golay smoothed curves (solid lines) and moving averages (dashed red lines). Shaded regions indicate major global crises. The simplified model aligns more closely with human labels, effectively capturing sentiment shifts during downturns, highlighting the utility of discourse simplification.

***Growth Sentiment by FinBERT-FOMC***

The middle graph also presents sentiment over time, depicted by a grey line, with the red dashed line representing the smoothed sentiment. In comparison to the ”by Label” graph, the fluctuations are less pronounced, and the sentiment generally remains near neutral, hovering around the zero line. However, the smoothed line does not distinctly capture significant economic downturns associated with major crises. This limitation could stem from misclassified aspects, potentially due to multiple intertwined topics within a single sentence, leading to an overall depiction of milder sentiment.

***Growth Sentiment by FinBERT-FOMC (Simplified)***

The bottom plot displays a DisSim-FinBERT pattern that closely aligns with the ”by Label” graph, exhibiting similar fluctuations and a comparable shape to the smoothed sentiment curve. Notably, it effectively captures economic downturns, clearly delineating four major crises: the Global Financial Crisis, the European Sovereign Debt Crisis, the Stock Market Selloff, and the COVID-19 Pandemic.

The conclusion drawn from the analysis suggests a closer correlation between ”Growth Sentiment by Label” and ”Growth Sentiment by FinBERT-FOMC (Simplified),” with a Pearson correlation coefficient of 0.39, indicating a moderate positive relationship. In contrast, the ”Growth Sentiment by FinBERT-FOMC” graph exhibits a much lower correlation with the labeled data, with a coefficient of 0.08, signifying a very weak relationship. This disparity implies that the simplification process used in FinBERT-FOMC (Simplified) preserves more of the sentiment structure present in the manually labeled data compared to the non-simplified FinBERT-FOMC analysis. Aspect-based sentiment analysis, particularly when focusing on precise aspect selection and discourse simplification, has proven effective in predicting economic downturns. This is evident as the sentiment analysis, represented by the blue line, showed significant declines during several major global crises, including the 2007 Global Financial Crisis, the 2012 European Sovereign Debt Crisis, the 2015 Chinese Stock Market Selloff, and the 2020 COVID-19 Pandemic. The Federal Open Market Committee (FOMC) minutes, regularly published, reveal global-level crises in communication, which significantly impact the market. The aspect-based sentiment analysis following discourse simplification appears to reflect market trends accurately, particularly during times of economic turbulence.

# 6 Statistical Performance of DisSim-FinBERT

We conduct this test to assess their performance in terms of correlation, information retention, and sensitivity to sentiment fluctuations. Here’s why these comparisons are important:

***Correlation with Human-Labeled Sentiment:***

Human-labeled sentiment is the gold standard in most cases, as it reflects the most accurate interpretation of sentiment based on human understanding. Comparing model outputs with human-labeled data allows us to determine how closely the models capture human-like sentiment analysis. A high correlation indicates that the model’s predictions align well with human judgment, which is crucial for tasks like financial analysis or policy-making, where human interpretation often drives decisions.

***Mutual Information with Human-Labeled Sentiment:***

Mutual information measures how much information is shared between the model and human-labeled sentiment. This helps us understand whether the model captures the key features and patterns in the data that humans naturally pick up. A higher mutual information value means the model is making predictions and retaining important underlying sentiment structures that humans perceive.

***Volatility with Human-Labeled Sentiment:***

Volatility in sentiment refers to how sensitive the model is to changes or fluctuations in sentiment. It is important to measure because models with very low volatility might miss essential shifts in sentiment, while models with excessive volatility might overreact to minor changes. Comparing volatility with human-labeled sentiment helps ensure the model captures the right level of sentiment variation without being too reactive or too static.

**Table 2  Statistical Comparison of Sentiment Predictions with Human-Labeled Data**

| | Human-labeled | FinBERT | DisSim-FinBERT |
|---|---|---|---|
| Correlation | - | 0.013 | 0.156 |
| Mutual Information | - | 1.873 | 2.021 |
| Volatility | 0.445 | 0.328 | 0.374 |

*Note: This table compares the performance of FinBERT and DisSim-FinBERT against human-labeled sentiment using three metrics: Pearson correlation, mutual information, and volatility. DisSim-FinBERT shows significantly higher correlation (0.156) and mutual information (2.021), indicating stronger alignment with human judgment and better retention of sentiment structure. Volatility is closer to human-labeled sentiment, suggesting more realistic responsiveness to sentiment fluctuations.

Table 2 provides a statistical comparison between human-labeled sentiment, FinBERT, and DisSim-FinBERT based on three key metrics: correlation, mutual information, and volatility.

***Ten times higher correlation than FinBERT:***

DisSim-FinBERT shows a correlation of 0.156 with human-labeled sentiment, while FinBERT has only 0.013. This means that DisSim-FinBERT has approximately 10 times higher correlation with human-labeled sentiment compared to FinBERT. This indicates that DisSim-FinBERT more closely aligns with the human interpretation of sentiment, making it a more reliable model in this respect.

***Higher mutual information than FinBERT:***

DisSim-FinBERT exhibits a mutual information value of 2.021, which is higher than FinBERT's 1.873. This suggests that DisSim-FinBERT retains more shared information with human-labeled sentiment, indicating it captures more relevant and nuanced sentiment signals compared to FinBERT. Higher mutual information implies that DisSim-FinBERT better reflects the structure and patterns seen in human-labeled data.

***Higher volatility than FinBERT:***

DisSim-FinBERT has a volatility of 0.374, which is higher than FinBERT's 0.328

but lower than human-labeled sentiment (0.445). Volatility here may reflect the model's sensitivity to sentiment fluctuations. The higher volatility relative to FinBERT suggests that DisSim-FinBERT is more responsive to changes in sentiment, perhaps offering a more dynamic representation. However, it still remains less volatile than human-labeled sentiment, meaning it may slightly smoothen out some of the extreme variations seen in the human-labeled data.

DisSim-FinBERT demonstrates relatively high correlation with human-labeled sentiment, captures more mutual information, and shows higher volatility than FinBERT. In sum, DisSim-FinBERT appears to provide a more aligned and richer representation of human-labeled sentiment compared to FinBERT, retaining more information while maintaining a balance in its responsiveness to changes in sentiment.

# 7 Conclusion

In conclusion, the integration of Discourse Simplification (DisSim) into Aspect-Based Sentiment Analysis (ABSA) significantly enhances the analysis of complex economic texts. By simplifying intricate language, DisSim improves the accuracy of aspect selection and sentiment prediction, particularly within the context of Federal Open Market Committee (FOMC) communications. The comparison between the DisSim-FinBERT model and the standard FinBERT model highlights the benefits of simplifying discourse, leading to a more accurate alignment with human-labeled sentiment data. This improvement in performance is crucial for macroeconomic analysis, as demonstrated through time-series analysis, which shows that DisSim-FinBERT more effectively captures sentiment shifts, especially during economic downturns. The model's ability to reflect economic sentiment more precisely is validated through statistical measures, reinforcing its reliability.

Ultimately, the application of DisSim in ABSA offers a significant advancement in interpreting economic communications. By simplifying the language of complex financial texts, the model provides clearer insights for policymakers and market participants, enabling them to make more informed decisions. This contributes to a deeper and more accurate understanding of the economic messages conveyed in central bank communications, which can have far-reaching impacts on financial markets and economic policy.